\documentclass[journal=nalefd,manuscript=letter]{achemso}
\usepackage[version=3]{mhchem} % Formula subscripts using \ce{}
\usepackage{amssymb}
\usepackage{amsmath}
\usepackage{natmove}

\author{L. D. Alegria}
\author{M. D. Schroer}
\author{A. Chatterjee}
\affiliation{Department of Physics, Princeton University, Princeton, New Jersey 08544}
\author{G. R. Poirier}
\affiliation{Princeton Institute for the Science and Technology of Materials, Princeton University, Princeton, New Jersey 08544}
\author{M. Pretko}
\author{S. \nolinebreak  K. \nolinebreak  Patel}
\author{J. R. Petta}
\email{petta@princeton.edu}
\affiliation{Department of Physics, Princeton University, Princeton, New Jersey 08544}
\affiliation{Princeton Institute for the Science and Technology of Materials, Princeton University, Princeton, New Jersey 08544}

\title{Structural and Electrical Characterization of \ce{Bi2Se3} Nanostructures Grown by Metalorganic Chemical Vapor Deposition}
\begin{document}
\begin{abstract}
We characterize nanostructures of \ce{Bi2Se3} that are grown via metalorganic chemical vapor deposition using the precursors diethyl selenium and trimethyl bismuth. By adjusting growth parameters, we obtain either single-crystalline ribbons up to $10$ $\mu\text{m}$ long or thin micron-sized platelets. Four-terminal resistance measurements yield a sample resistivity of 4 m$\Omega$-cm. We observe weak anti-localization and extract a phase coherence length $l_\phi$ = 178 nm and spin-orbit length $l_\text{so}$ = 93 nm at $T$ = 0.29 K. Our results are consistent with previous measurements on exfoliated samples and samples grown via physical vapor deposition.
\end{abstract}


\ce{Bi2Se3} is a strong topological insulator (TI) with a single Dirac cone and chiral spin texture \cite{Hasan2010,Qi2010a}. Surface-sensitive probes have directly accessed the topological surface states, but transport measurements have proven difficult due to bulk contributions to the conductivity \cite{Xia2009,Roushan2009}. Efforts to isolate surface transport properties include mechanical exfoliation, electrical gating, and chemical doping \cite{Hor2009, Analytis2010, Checkelsky2009, Checkelsky2011, Steinberg2011, Kim2012}. Another way to reduce bulk conduction is to directly synthesize nanostructures with a large surface-to-volume ratio \cite{Peng2010,Xiu2011,Hong2012}. We demonstrate the controlled synthesis of \ce{Bi2Se3} nanoribbons using metal-organic chemical vapor deposition (MOCVD), a standard process for rational nanostructure synthesis, and we characterize the nanoribbons using electron microscopy and low-temperature magnetotransport measurements.

\ce{Bi2Se3} has been the focus of many TI experiments due to its comparatively large bulk band gap of 0.35 eV and simple surface band structure \cite{Xia2009}.  \ce{Bi2Se3} adopts a rhombohedral structure belonging to the space group $D^5_{3d}$ ($R3\bar{m}$) \cite{PerezVicente1999}. The crystal consists of 2D hexagonal lattices of either Se or Bi, which stack according to the sequence Se-Bi-Se-Bi-Se. Adjacent layers of Se are Van der Waals bonded, such that \ce{Bi2Se3} preferentially forms sheet-like structures.  In addition to TI research, \ce{Bi2Se3} and related compounds are of interest for high performance thermoelectric materials \cite{Dresselhaus2007}. In that context, various nanostructures, including quintuple layer nanotubes, have been synthesized by co-reduction from solution, template-assisted electrodeposition, and solid-source vapor transport \cite{Cui2004,Jagminas2008,Lee2008}.  Many of these convenient methods have subsequently been employed for TI experiments \cite{Kong2010a, Peng2010}. However, factors including sample purity, uniformity of growth, and precise control over the Se/Bi flux ratio favor molecular beam epitaxy (MBE) and MOCVD.

Previous MOCVD studies of \ce{Bi2Se3} have focused on the development of appropriate precursors and growth conditions for thin film growth\cite{AlBayaz2002}.  In one case the formation of \ce{Bi2Se3} platelets alongside BiPO$_4$ nanowires was observed in the decomposition of the oxygen and phosphorus-containing bismuth complex Bi[Se$_2$P(O$i$Pr)$_2$]$_3$\cite{Lin2007}.  Here we pursue the MOCVD synthesis of \ce{Bi2Se3} nanostructures using separate Se and Bi precursors, allowing individual control of the precursor partial pressures.

We grow nanostructures in a laboratory-scale MOCVD reactor, similar to one used to produce high quality InAs nanowires \cite{Schroer2010a,Schroer2010b}. Mass flow controllers admit fixed flows of H$_2$ carrier gas through two bubblers containing the liquid metal-organic precursors trimethyl bismuth (TMBi) and diethyl selenium (DESe) \cite{Source}. The precursor vapors and H$_2$ gas flow into a cold-walled chamber and impinge on a heated sample holder containing a Si (100) substrate that is prepared with a 5 nm thick Au seed layer.

A uniform layer of \ce{Bi2Se3} nanostructures forms on the $\sim$1 cm$^2$ substrate under a range of growth conditions. We generally obtain two different types of structures:  nanoribbons with 10 -- 30 nm thickness and lengths of several microns, or nanoplates as thin as 10 nm with roughly 1 $\mu$m lateral dimensions. The growth temperature, $T_{\rm g}$, and precursor partial pressure ratio, $r$ = $p_\text{DESe}/p_\text{TMBi}$, determine which type of structures we obtain. Figure 1 shows scanning electron microscope (SEM) images of samples obtained for a range of growth parameters. High-yield growth occurs for $T_{\rm g}$ = 470 $^\circ$C, with a chamber pressure $P$ = 100 Torr, carrier gas flow of 600 sccm $\text{H}_2$, TMBi partial pressure $p_{\text{TMBi}}$ = 1 $\times$ $10^{-5}$ atm, and $r$ = 30. Typical growth times are 15 minutes. In order to systematically study changes in the growth morphology with reaction conditions, we take this parameter set as a reference from which we vary individual parameters. Under these conditions, nanoribbon growth begins once $r$ exceeds $\sim$ 7. Between $r$ = 7 -- 33, the reaction products transition from narrow ribbons to wide ribbons and plates, accompanied by an increase in density.  Little change in morphology is observed above $r$ = 33 which suggests that Bi limits the growth at such high precursor ratios.

\begin{figure}
		\includegraphics{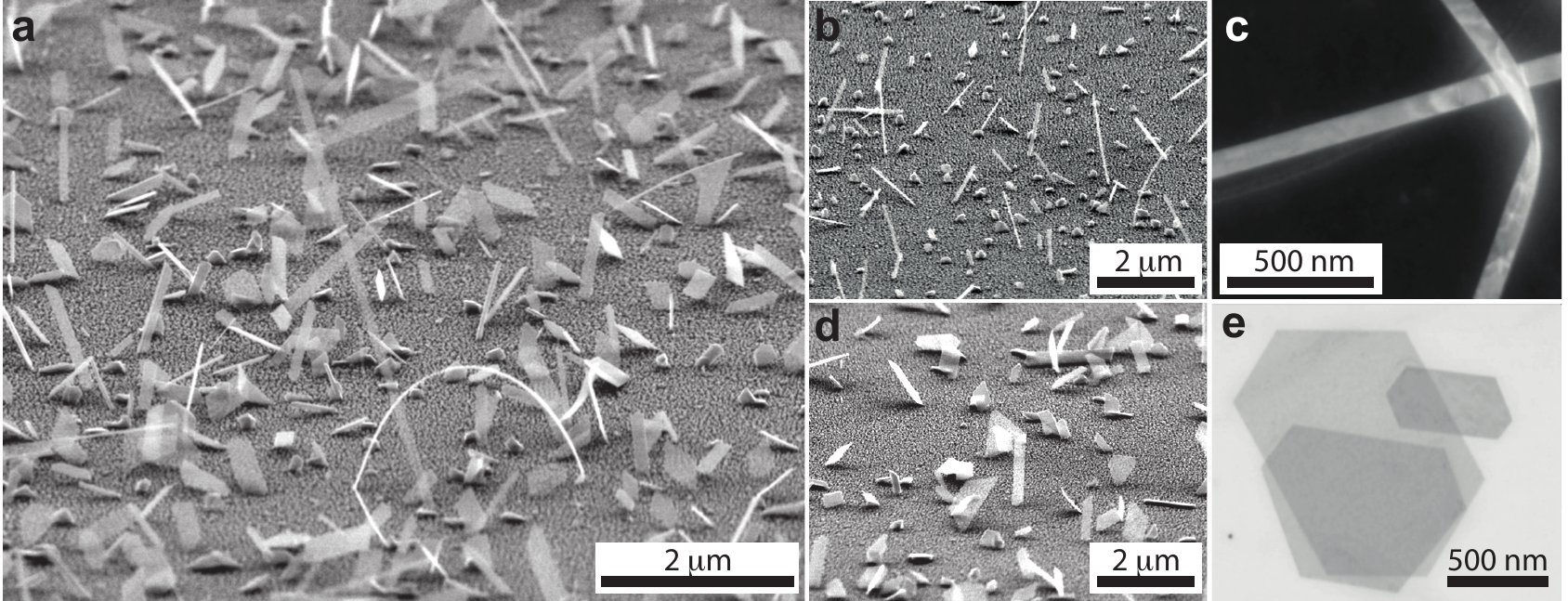}
\caption{\label{fig1} \ce{Bi2Se3} nanoribbons and platelets: (a,b,d) SEM images of as-grown samples, and (c,e) images obtained after deposition on TEM grids. (a) Diverse growth is obtained from a 15 minute growth run on a Si (100) substrate with 5 nm Au seed layer, with $T_{\rm g}$ = 470 $^\circ$C, $P$ = 100 Torr, 600 $\text{sccm}$ $\text{H}_2$ carrier gas flow, TMBi partial pressure $p_{\text{TMBi}}$ = 1 $\times$ $10^{-5}$ atm, and precursor ratio $r$ = 33.  (b) A reduced precursor ratio $r$ = 12 results in narrow nanoribbons of comparatively well-defined widths $70\pm20$ nm.  (c) Dark field scanning transmission electron microscope (STEM) image of two nanoribbons with $85\times10$ $\text{nm}^2$ cross section.  The nanoribbons are single crystal and show stress induced fringes. (d) We obtain platelets at $r$ = 30 and an elevated growth temperature $T_{\rm g}$ = 490 $^\circ$C.  (e) $\sim1$ $\mu\text{m}^2$ platelets imaged by STEM.}
\end{figure}

We also consider the temperature and time dependence of the growth. With  all other growth parameters fixed as above, nanoribbons are obtained at a growth temperature $T_{\rm g}$ = 470 $^\circ$C and widen into plates above 480 $^\circ$C (see Figure 1d).  Longer growth runs of 30 minutes produced ribbons up to 10 $\mu \text{m}$ long, maintaining cross sections on the order of $10\times100$ $\text{nm}^2$.

In general, epitaxial nanostructure growth can occur by several mechanisms, including vapor-liquid-solid (VLS) and vapor-solid (VS) growth \cite{Kolasinski2006,Mohammad2010}.  In typical VLS growth, a metal droplet nucleates crystal growth and determines the nanowire diameter.  \ce{Bi2Se3} nanostructure growth has been demonstrated with and without seed particle involvement\cite{Kong2010a,Cha2012}.  Under the above growth conditions, we find that growth occurs when the Si substrate is prepared with a 5 nm film of Au, but not on bare Si.  We do not observe a gold particle at the free end of the nanoribbon, as seen in the solid-source growth method described by Kong et al. \cite{Kong2010a}.  Instead, many ribbons clearly originate from gold nanoparticles on the substrate.  VLS growth in our MOCVD growth process may therefore be occurring at the base of the nanowire via root catalyzed growth, as has been observed in other materials \cite{Kolasinski2006}.

We establish that the samples are single-crystalline \ce{Bi2Se3} using transmission electron microscopy (TEM).  The nanostructures are first freed from the growth substrate by sonication in ethanol and then transferred to porous carbon TEM grids for imaging in a Phillips CM200 transmission electron microscope.  Figure 2 displays a high resolution TEM analysis of a typical nanoribbon.   By performing selected-area electron diffraction at various points along the nanoribbon, we confirm the single-crystal, rhombohedral structure of the ribbons.

\begin{figure}
\includegraphics{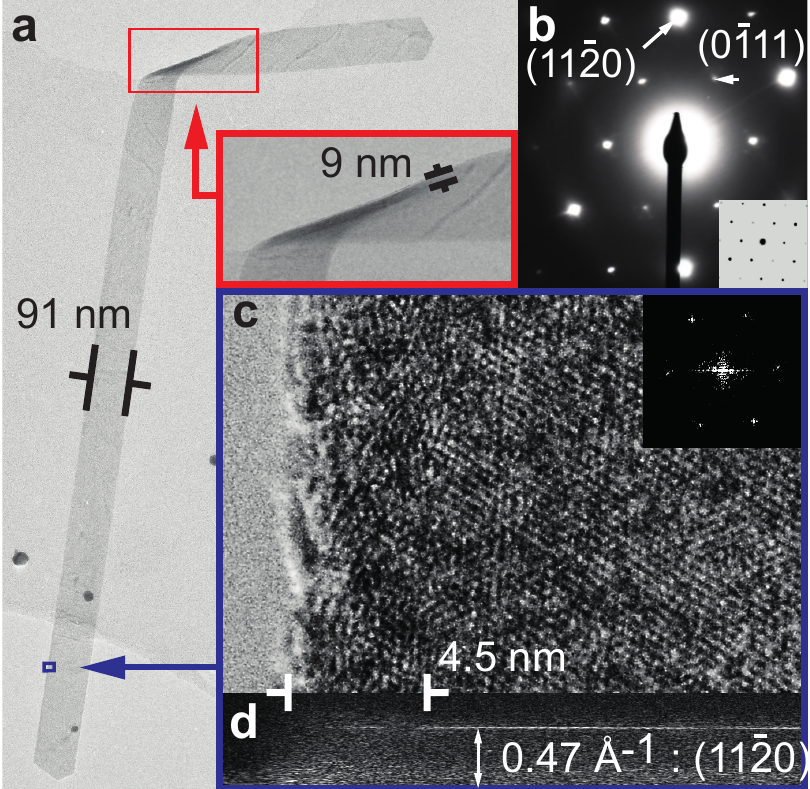}
\caption{\label{fig2} (a) TEM image of a 2 $\mu$m long nanoribbon folding around the pore of a TEM grid. Growth conditions for this sample were: $T_{\rm g}$ = 480 $^\circ$C, $P$ = 100 $\text{Torr}$, $t$ = 30 $\text{min}$, 600 $\text{sccm}$ flow of $\text{H}_2$, $p_{\text{TMBi}} =1\times10^{-5}$ atm, $r$ = 30. A fold in the nanoribbon reveals a thickness of 9 $\text{nm}$. (b) TEM electron diffraction pattern. The inset shows the expected diffraction pattern. (c) A HRTEM image shows that the nanoribbon is single crystal (inset: Fourier transform obtained from the real space image). (d) The FFT of the columns of the image in (c) shows a peak at 0.47 $\text{\AA}^{-1}$, consistent with the (11$\bar{2}$0) growth direction. The peak fades at the edge of the nanoribbon, indicating an amorphous region with width $\sim$ 4.5 nm, typical of atmospherically exposed samples.}
\end{figure}

The nanoribbons grow in the (11$\bar{2}$0) direction with lattice constant $a=4.1\ \text{\AA}$, consistent with the spacing expected from the bulk crystal structure.  TEM-based energy dispersive X-ray spectroscopy (EDS) indicates a 2:3 ratio of Bi and Se in the nanoribbons to within the accuracy of the measurement.  The samples are exposed to atmosphere after growth and exhibit an amorphous edge region several nanometers wide.  The upper and lower surfaces, which lack the dangling bonds as on the edges, must have considerably less irregularity, or crystallinity would not be observed in samples as thin as in Figure 2.

A large fraction of the crystallites shown in Figure 1a are partially transparent under the SEM (at 30 keV), indicating the formation of very thin, suspended nanoplates.  We use TEM and AFM to quantitatively extract the sample thickness. By directly imaging the thicknesses of wires bent around the pores of the TEM grids (as in Figure 2) we find an average thickness of $\sim10$ nm.  An AFM study of 18 nanoribbons gives thicknesses $20\pm10$ nm.  The discrepancy is probably due to an unintentional thickness bias imposed during sample preparation for the two imaging methods.  The nanoribbon thicknesses lie below the distribution of dimensions produced in the solid-source method referred to above. We find that the solid-source growth method typically produces ribbons 30--100 nm in thickness, consistent with other reports in the literature \cite{Kong2010a}.

\begin{figure}
\includegraphics{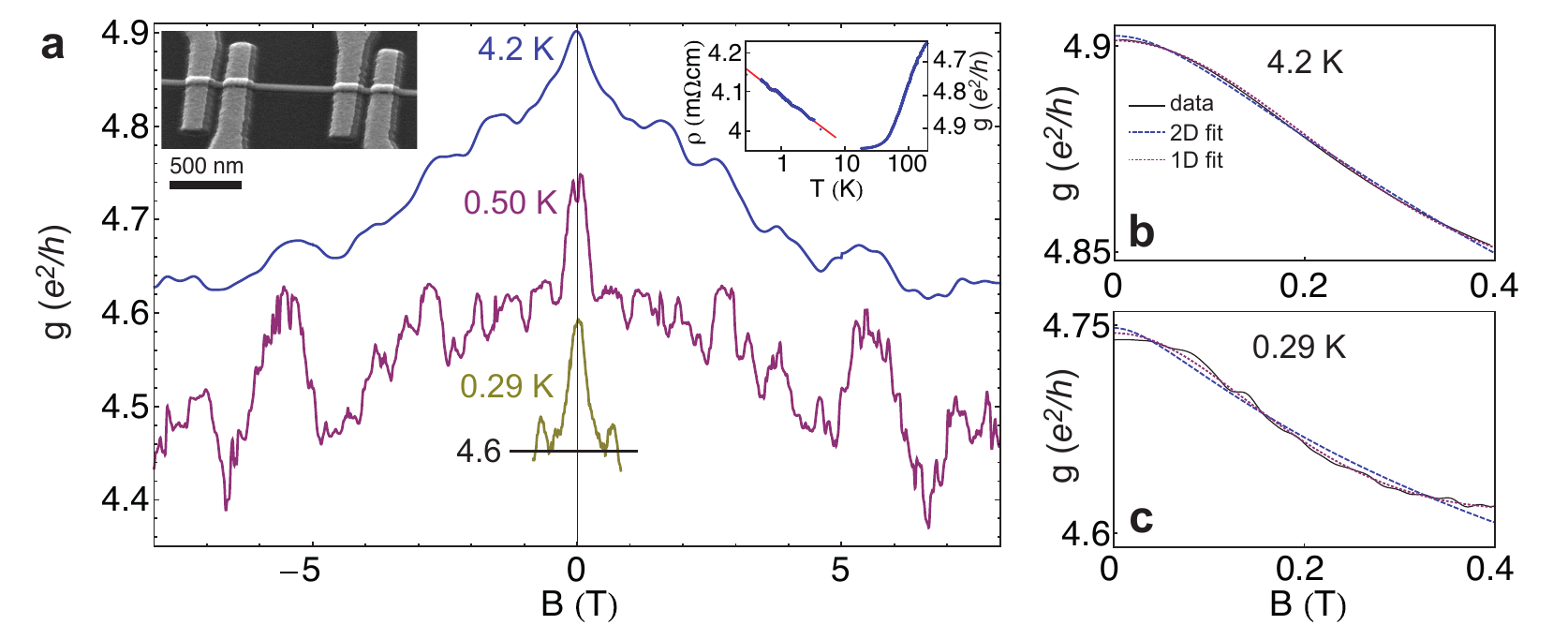}
\caption{\label{fig3}  (a) The four probe conductance of a nanoribbon in perpendicular magnetic field shows weak anti-localization and universal conductance fluctuations. The trace taken at 0.29 K is offset for clarity. Left inset: SEM image of a typical device. Right inset: Sample resistivity as a function of temperature shows a metallic dependence above 10 K, typical of \ce{Bi2Se3} samples. (b--c) Low field data taken at 4.2 K and 0.29 K are fit to 2D (ribbon width greater than the coherence length) and 1D (ribbon width less than the coherence length) models of weak anti-localization.}
\end{figure}

Ribbons are further characterized by low-temperature magnetotransport measurements, which are summarized in Figure 3.  Four probe devices are made by eliminating the native oxide in e-beam defined contact areas using a low-energy ion etch, followed by thermal evaporation of Ti/Au contacts.  We have studied a nanoribbon with thickness $17\pm5$ nm, width $W$ = $170\pm5$ nm, and length $L$ = $380\pm10$ nm, where dimensions were measured by SEM after performing transport measurements.  Using these dimensions, the calculated resistivity is $4\pm1$ m$\Omega$-cm, similar to that of bulk and nanoribbon \ce{Bi2Se3} samples with $n$ = $10^{18}$ cm$^{-3}$. The resistance versus temperature profile is also similar to such samples, having a metallic profile at high temperatures and a $\log(T)$ dependence below 10 K \cite{Checkelsky2009, Analytis2010, Hong2012}. The temperature dependence is consistent with weak anti-localization (WAL) in \ce{Bi2Se3}, which we further explore by measuring the magnetoconductance.

Perpendicular magnetic field sweeps at 4.2 K, 0.50 K, and 0.29 K show reproducible, aperiodic fluctuations consistent with universal conductance fluctuations, and a low field conductance peak consistent with WAL in \ce{Bi2Se3} \cite{Checkelsky2011}. The presence of conductance fluctuations indicates that the coherence length is on the same order as the sample dimensions.  We fit the low field data ($|B|$ < 0.4 T) at 4.2 K and 0.29 K to the 2D WAL model, which is valid in the limit of $W\gg l_\phi \gg l_\text{so}$,

\begin{equation}
\Delta g =  \alpha \frac{e^2}{ \pi h} \left[ ln \left(\frac{B_\phi}{B}\right) - \Psi \left(\frac{1}{2}+\frac{B_\phi}{2 B}\right) \right]
\end{equation}

where $\Delta g$ = $g(B)$ - $g(B=0)$, $\Psi$ is the digamma function, $h$ is Planck's constant, $e$ is the elementary charge, $\alpha$=1/2 for a single 2D channel, and the phase breaking field $B_\phi$ = $h/4 e l_\phi^2$ is defined in terms of the coherence length $l_\phi$ \cite{Hikami1980}.  We take $\alpha$ and $l_\phi$ as fit parameters.  At 4.2 K and 0.29 K, we obtain $\alpha$ = 0.41 and 0.43 respectively, consistent with a single coherent conduction channel \cite{Steinberg2011}.  At 4.2 K we find $l_\phi$ = 119 nm, but at 0.29 K we find $l_\phi$ = 192 nm which is greater than the channel width, invalidating the 2D treatment at that temperature.  We therefore also fit according to a diffusive 1D model, which is valid for $l_\phi \gg W$ \cite{Santhanam1984,Roulleau2010},

\begin{equation}
\Delta g =  - \frac{2 e^2}{h L} \left[\frac{3}{2} \left(\frac{1}{l_\phi^2} + \frac{4}{3}\frac{1}{l_\text{so}^2} + \frac{1}{3}\left(\frac{ e W B}{h}\right)^2\right)^{-\frac{1}{2}}  - \frac{1}{2}\left(\frac{1}{l_\phi^2} + \frac{1}{3}\left(\frac{ e W B}{h}\right)^2\right)^{-\frac{1}{2}} \right].
\end{equation}

Taking $l_\text{so}$ and $l_\phi$ as fit parameters, Eqn.\ (2) provides a marginally better fit to the WAL peak at 4.2 K (yielding $l_\phi$ = 113 nm and $l_\text{so}$ = 69 nm) and a visibly better fit at 0.29 K, where we obtain $l_\phi$ = 178 nm and $l_\text{so}$ = 93 nm. The consistency of the two treatments suggests that the coherence length crosses over between the 2D and 1D limits in this temperature range. The observed short spin-orbit length likely arises from a combination of transport through the surface states and bulk states that are spin-split by surface Rashba fields, and is comparable in magnitude to other strong spin-orbit nanomaterials such as InAs nanowires \cite{King2011,Hansen2005}.

In summary, we demonstrate the synthesis of \ce{Bi2Se3} nanostructures using MOCVD. MOCVD allows independent control of the Bi and Se concentration during growth. High resolution electron microscopy confirms that the samples exhibit a high degree of structural order. Magneto-transport measurements yield a spin-orbit length $l_\text{so}$ $\sim$ 100 nm, and phase coherence length $l_\phi$ = 100--200 nm. Four-probe measurements give a resistivity
of 4 m$\Omega$-cm, indicating that the doping levels are comparable to \ce{Bi2Se3} samples grown using other methods. The measured phase coherence lengths are also similar to samples fabricated by other means \cite{Cha2012, Checkelsky2011}. Our results indicate that MOCVD samples are of sufficiently high quality to enable transport based studies. Further work is required to clarify and refine the growth mechanism, and to determine whether similar MOCVD processes are possible in more sophisticated TI materials such as $Bi_2Te_2Se$ \cite{Ren2010}.

\acknowledgement
We thank Bob Cava and Phuan Ong for useful discussions, Sian Dutton, Minkyung Jung, Sunanda Koduvayur Parthasarathy, Chris Quintana, and Jian Zhang for technical contributions, and Nan Yao at the Princeton Imaging and Analysis Center for assistance characterizing samples. Research was supported by the Sloan and Packard Foundations, and the NSF funded Princeton Center for Complex Materials, DMR-0819860. We acknowledge the use of the PRISM Imaging and Analysis Center, which is supported in part by the NSF MRSEC program.

\end{document}